# MALİYET DÜŞÜRME YÖNTEMİ OLARAK DİJİTAL İKİZ[*]

Prof. Dr. Süleyman YÜKÇÜ[**]

Dr. Ömer AYDIN[***]

**ÖZ**

Sensörler, endüstri 4.0, nesnelerin interneti, makine öğrenmesi ve yapay zeka gibi kavramların son yıllarda hayatımıza girmesi ile birçok alan bu kavramlardan etkilenmiştir. Siber fiziksel sistemlerin bu kavramlar ile etkileşimi sonucunda dijital ikiz kavramı ortaya çıkmıştır. Dijital ikiz kavramı, ortaya çıkması ile birlikte birçok alanda kullanılmaya başlanmıştır. Bu modelin kullanımı özellikle karar verme süreçlerinde ciddi kazanımlar sağlamıştır. Karar verme süreçlerindeki kazanımlar her alanda katkı sağladığı gibi maliyet anlamında da değişikliklere neden olmaktadır. Bu çalışmada dijital ikiz kavramının tarihsel gelişimine değinilmiş, dijital ikizin kullanım alanları hakkında genel bilgiler verilmiştir. Bu bilgiler ışığında dijital ikiz modelinin maliyet etkisi, dolayısı ile maliyet muhasebesi penceresinden görünümü ve maliyet düşürme yöntemi olarak kullanılması değerlendirilmiştir. Türkçe literatürde yeteri kadar kaynak bulunmaması ve maliyet muhasebesi bakış açısı ile çalışmalara ışık tutmak adına bu çalışma yapılmıştır.

**Anahtar Kelimeler:** Dijital İkiz, Maliyet Muhasebesi, Siber Fiziksel Sistemler, Yapay Zeka, Nesnelerin İnterneti, Sensör

**JEL Sınıflandırması:** C80, L86, M41

## DIGITAL TWIN AS A COST REDUCTION METHOD

**ABSTRACT**

Many fields have been affected by the introduction of concepts such as sensors, industry 4.0, internet of things, machine learning and artificial intelligence in recent years. As a result of the interaction of cyber physical

---

systems with these concepts, digital twin model has emerged. The concept of digital twin has been used in many areas with its emergence. The use of this model has made significant gains, especially in decision making processes. The gains in decision making processes contribute to every field and cause changes in terms of cost. In this study, the historical development of the concept of digital twin has been mentioned and general information about the usage areas of digital twin has been given. In the light of this information, the cost effect of the digital twin model, therefore its appearance from the cost accounting window and its use as a cost reduction method were evaluated. This study was carried out in order to shed light on the studies with the insufficient resources in the Turkish literature and the cost accounting perspective.

## 1. GİRİŞ

Endüstri mühendisliği alanında yaygın uygulamaları bulunan siber fiziksel sistemlerin (CPS) operasyonel modellerine nesnelerin interneti (IoT), makine öğrenmesi ve yapa zeka (AI) teknolojilerinin evrimi ile entegrasyonunun sağlanması neticesinde dijital ikiz (DT) kavramının uygulanabilirliği kolaylaşmıştır. Fiziksel süreçlerin sanal ortamda sürekli algılanması ve buradan elde edilen veriler ile kararlar almak ve eylemler gerçekleştirmek temel zorluklar olarak görülmektedir. Siber fiziksel sistemler imalat, sağlık ve tüketici hizmetleri, enerji sistemleri gibi birçok alanda kullanılmaktadır (Koulamas ve Kalogeras 2018).

Dijital ikiz, fiziksel bir nesneye karşılık gelen ve fiziksel nesnenin mevcut durumunu tam ve doğru bir şekilde yansıtan dijital kopyasıdır. Başka bir ifade ile dijital ikiz, bir işlem, ürün veya hizmetin sanal bir modelidir. Dijital ikize Nesnelerin İnterneti (Endüstri 4.0), yapay zeka ve yazılım çözümleri eklenmiştir. Gerçek zamanlı olarak fiziksel ortamdan veri toplamak için ise duyargalar (sensör) kullanılmaktadır. Duyargalar aracılığı ile fiziksel ortamdan toplanan veriler bilgisayar ortamına aktarılır. Aktarılan gerçek zamanlı veriler ile fiziksel bir sürecin, ürünün veya hizmetin bilgisayar ortamında sanal bir ikizi oluşturulur. Bu süreç modern mühendislikte yenilik ve performansı arttırmak için kullanılan temel yöntemlerden biri haline gelmektedir. Son derece gelişmiş analiz, izleme ve çıkarım yapma gibi yeteneklere sahip olan dijital ikiz sistemleri birçok şirket tarafından kullanılmaya başlanmıştır.

Dijital ikiz modeli 2002 yılından bu yana var olmasına karşın bu terim 2010 yılında gün yüzüne çıkmıştır. Bu tarihten sonra ürün yaşam döngüsü boyunca sunması öngörülen birçok fayda nedeni ile gün geçtikçe popülerliği artmıştır. Dijital ikiz kavramı şu an dünyanın en büyük şirketleri tarafından büyük ilgi görmekte ve geliştirilmeye devam etmektedir. Dijital ikiz, kullanım alanını ve yaygınlığını günden güne arttırsa da birçok kişi ve işletme için yeni bir kavramdır. Bu kavram, 2000'li yılların başında ortaya koyulan "akıllı ürünler" kavramının evrimleşmesi ile ortaya çıktığı söylenebilir (Kiritsis 2011).

Dijital ikiz fiziksel ve dijital dünya arasında bir köprü olarak düşünülmelidir. Dijital ikiz sistemlerinde, gerçek zamanlı durum, çalışma koşulu veya konum gibi verileri toplamak için duyargaları kullanan akıllı bileşenler fiziksel bir nesneye entegre edilir. Bu akıllı ve çevrimiçi çalışan bileşenler, duyargaların izlediği tüm verileri alan ve işleyen bulut tabanlı bir sisteme bağlanırlar. Burada toplanan veriler çeşitli ölçüt ve verilere göre analiz edilir. Bu sistem vasıtası ile sanal bir ortamda, fiziksel dünya uygulanabilecek işinizi dönüştürmek, çeşitli testler ve analizler yapmak gibi birçok işlem uygulanabilir hale gelmektedir.

Üretim biriminin elektronik kopyası olan dijital ikizde üretim birimlerinde yapabilecek her türlü iyileştirme yapılabilir. İyileştirme süreçlerin veya makinelerin değiştirilmesi, iyileştirilmesi gibi radikal değişiklik gerektirebilir. Fiziksel değişiklik her zaman zor, zaman alıcı ve pahalıdır. Bu değişiklikleri dijital ikizde yapmak kolay ve daha ucuzdur.

Günümüz işletmeleri kıran kırana rekabet ortamında yarışırken maliyet düşürme yaklaşımı en kontrol edilebilen işletme amacıdır. Çünkü birçok etmen kontrol edilememektedir. Örneğin fiyat işletmenin kontrolü dışında serbest piyasada oluşmaktadır.

Hedef maliyetleme, bir pazar payına ulaşabilmek için kullanılan ve satış fiyatına göre hesaplanan, pazar bazlı maliyeti ifade etmektedir (Yükçü ve Gönen 2011, 75). Hedef maliyetleme yaklaşımında, mamulun tasarlanıp daha sonra kaça mal olduğunu öğrenmek yerine, bir hedef maliyet belirlenerek mamulun ona göre tasarlanmasını ve böylece hedeflenen maliyete ulaşılması amaçlanmaktadır. Bir başka deyişle hedef maliyetlemenin amacı, kar ve maliyet planlamasını birlikte yürüterek, uygun karlar elde edilmesini sağlamaktır (Erden 2003, 87-88).

Bu çalışmanın amacı, maliyet odaklı çalışan işletmelerde önemli bir maliyet düşürme yöntemi olarak görülen dijital ikizi tanıtmak, dijital ikiz ile maliyet düşürmenin yöntemleri konusunda önerilerde bulunmaktır.

## 2. DİJİTAL İKİZ KAVRAMI

Günümüz dünyasında gerçek zamanlı veri kullanımı ve yaygınlığı artmaktadır. Bu verilerin işlenmesi, öğrenme tekniklerinin kullanılması ve yapay zeka ile kararlar alınması fiziksel ürünlerin veya süreçlerin sanal ortamda temsil edilmesini anlamlı ve kolay kılmaya başlamıştır. Bu temsil edebilme imkanı ile fiziksel nesne veya süreçlerin birer kopyasının oluşturulması mümkün hale gelmiştir. Bu kopya dijital ikiz olarak adlandırılmaktadır. Dijital ikiz için belirlenmiş benzersiz ve küresel olarak kabul görmüş bir tanım bulunmamaktadır. Sadece yapılmış tanımlarda kabul görmüş belli noktalar bulunmaktadır. Öncelikle dijital ikiz sanal yani sayısal bir olgudur. Hem durağan hem de hareketli bölümleri içinde barındırmaktadır. Verinin elde edilmesi, iletilmesi ve sistemin simülasyonunun oluşturulmasını kapsamaktadır. Ürünün kendisi veya ürün döngüsündeki her örnek ve veriyi ele alır. Aslına bakılırsa mühendislik uygulamalarının birçoğunda gerçek dünyadaki bir nesnenin veya sürecin simülasyonu kullanıla gelmektedir. Dijital ikiz kavramı dijital ve fiziksel uzayların çift yönlü olarak entegrasyonudur (Koulamas ve Kalogeras 2018).

## 3. DİJİTAL İKİZ TARİHÇESİ

Dijital ikiz terimi ilk kez Shafto ve arkadaşları tarafından dile getirilmiştir (Shafto ve diğerleri 2010; Shafto ve diğerleri 2012; Schroeder ve diğerleri 2016, 13). Bununla birlikte, dijital ikiz ile çok benzer fikirler 2000'li yılların başında, örneğin Grieves tarafından geliştirilmiştir (Grieves ve Vickers 2017, 93; Främling ve diğerleri 2003). Bu dönemdeki kavramlar bir dijital ikizin mevcut vizyonuna benzer bazı özellikleri paylaşır, ancak henüz dijital ikizler olarak adlandırılamazlar.

Brussel ve diğerleri, Holonik imalat sistemi (HMS) için özerk ajanlar olan "holon" terimini ortaya atmışlardır (Brussel ve diğerleri 1999). Holonlar ortak bir hedefe ulaşmak için işbirliği yapabilir, rahatsızlıklara tepki verebilir ve süreci optimize edebilir. HMS üç temel yapı taşından oluşur (Koestler 1970):

1) Ürün yaşam döngüsü ve malzeme listesi hakkında güncel bilgiler gibi ürünün kendisi hakkında bilgi içeren ürün holonu,

2) Fiziksel bölümden oluşan kaynak tahsisi için kullanılan kaynak holonu (kaynak) ve bilgi işleme bölümü,

3) "İşi doğru ve zamanında yapmaktan sorumlu aktif bir varlık" olan sipariş holonu.

Wong ve diğerleri akıllı ürün kavramını tanıtmış ve ürünün yaşam döngüsü üzerindeki etkilerini incelemiştir (Wong ve diğerleri 2002). Akıllı bir ürünü aşağıdaki özelliklerden en azından bazılarına sahip olarak tanımlarlar:

1) Benzersiz kimlik.

2) Çevresi ile iletişim kurabilme.

3) Kendisi hakkında veri tutabilir veya saklayabilir.

4) Özelliklerini ifade edebilir, üretim gereksinimleri vb.

5) Kendi kaderine katılma veya kendi kaderine uygun kararlar verme yeteneği.

Yazılım ajanları, akıllı ürünlerin yukarıda belirtilen 4 ve 5 özelliklerini ele almasını sağlar. Wong ve arkadaşları yazılım ajanını "bağımsız bir şekilde akıl yürütebilen ve diğer ajanlar ve çevresi tarafından ortaya çıkan değişime tepki verebilen ve diğer ajanlarla işbirliği yapabilen farklı bir yazılım süreci" olarak tanımlar. Şekil 1, etiketli bir nesne örneğini göstermektedir. Etiket, ürünün kendisi ve yazılım ajanı hakkındaki bilgileri bir ağ üzerinden iletmek için kullanılır (Wong ve diğerleri 2002).

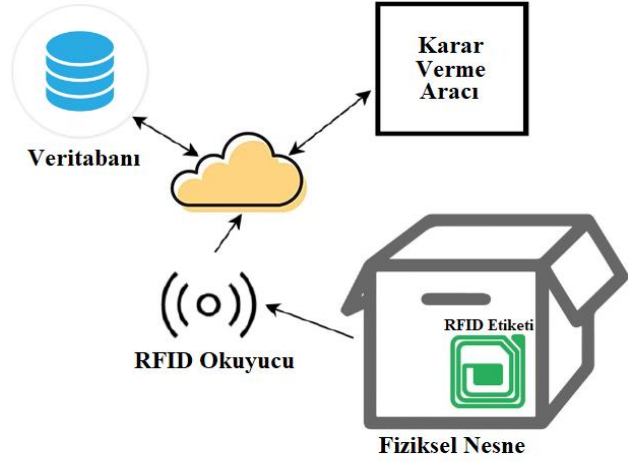

**Şekil 1. Akıllı ürün bir RFID etiketi ile tanımlanır ve böylece kendisi ve yazılım ajanı hakkındaki bilgiler eşleştirilebilir. Dijital ikiz kavramında, fiziksel ve dijital ikiz arasında çift yönlü bir iletişim vardır. (Wong ve diğerleri 2002, 2)'den tekrar çizildi.**

Hribernik ve diğerleri (2005), Wong ve arkadaşları (2002) tarafından ortaya atılmış ve yukarıda verilen ifadeyi "Ürün Avatarı" kavramının özelliklerini tanımlamada kullanmıştır. Her ürünün sanal gerçeklikte avatar adında bir dijital karşılığı vardır. Avatar otonom karar verme yeteneğine sahip kendine özgü bir nesnedir. Hribernik ve diğerleri ürün ile ilgili bilgileri yönetmek için Ürün Merkezli Yaklaşımı önermişlerdir (Hribernik ve diğerleri 2005; Hribernik ve diğerleri 2006). Geleneksel yaklaşımda bilgiler tek tek taraflarca saklanır ve bu nedenle kolay erişilebilir değildir. Ürün yaşam döngüsü sırasında toplanan verilere erişim; çalıştırma, bakım ve onarım gibi evrelerin optimizasyonuna olanak tanır.

Främling ve diğerleri bir ürünün tüm yaşam döngüsü boyunca bilgilerini yönetmek için ajan tabanlı bir mimari önermiştir (Främling ve diğerleri 2003). Her ürün belli bir amaç için çalışan, diğer ajanlarla iletişim kurabilen otonom bir yazılım bileşenine sahip ve "sanal karşılık" (Främling ve diğerleri 2003, 5) olarak adlandırılan bir ajana sahiptir (Wooldridge ve Jennings 1995; Holmström ve diğerleri 2002, 41). Bir ürünün ajanına internet üzerinden erişilebilir ve bu durum ürünün bilgisini erişilebilir kılar (Främling ve diğerleri 2003).

Grieves, gerçek alan, sanal alan ve bu alanlar arasındaki bağlantıdan oluşan Yansıtılmış Mekanlar Modeli'ni (Mirrored Spaces Model -MSM) tanıttı (Grieves, 2005). Sanal alandaki nesneler gerçek alandaki fiziksel muadillerine bağlanır ve durumlarını yansıtır. MSM, ürün verilerini kullanım ömrü boyunca erişilebilir kılarak ürün yaşam döngüsü yönetimine olanak tanır. MSM ilk olarak Bilgi Yansıtma Modeli (Information Mirroring Model- IMM) ve daha sonra dijital ikiz (Grieves ve Vickers 2017, 93-94) olarak yeniden adlandırıldı.

Grieves (2011) kitabında Ürün Yaşam Döngüsü Yönetimine (Product Lifecycle Management-PLM) sanal ürünlerin değerini sunmuş ve sonra IMM konseptini geliştirmiştir (Grieves ve Vickers 2017, 93). Grieves genel olarak, dijital ikiz kavramının yaratıcısı olarak kabul edilir, ancak dijital ikiz terimini ilk olarak Shafto ve arkadaşları kullanmıştır (Shafto ve diğerleri 2010).

Kiritsis akıllı ürünü "çeşitli zeka seviyelerinde algılama, bellek, veri işleme ve iletişim yeteneklerini içeren ürün sistemi" olarak tanımlamaktadır (Kiritsis 2011). Bir ürünün zekası dört seviyeye ayrılmıştır: Zeka seviyesi 1'deki ürün herhangi bir zeka içermez diğer yandan seviye 4'teki ürün karar verme ve çevresi ile iletişim kurma yeteneğine sahiptir. Akıllı ürünler, Ürün Yaşam Döngüsünü sonlandıran verilerin toplanmasını sağlar. Bu, üreticilerin kullanımında olan gerçek üründen veri almasını ve bakım işlemlerini iyileştirmesini sağlar, çünkü ürünün güncel durumu bilinmektedir. Akıllı ürünler ürün türünü odak noktası olmaktan çıkararak ürünü odak haline getirirler.

Siber-fiziksel Sistem (Cyber-Physical System - CPS) kavramı dijital ikiz kavramı ile yakından ilişkilidir, çünkü dijital ikiz "Siber-Fiziksel Sistemin siber kısmı" olarak görülebilir (Autiosalo 2018, 243). Lee siber-fiziksel sistemleri "hesaplamanın fiziksel süreçlerle bütünleşmesi" olarak tanımlamaktadır (Lee 2008, 363). Bir Siber-Fiziksel Sistemde, fiziksel verileri kontrol etmek için sensör verileri toplanır ve analiz edilir (Alam ve El Saddik 2017, 2050-2051). Siber-Fiziksel Sistemler dijital bir ikizin gelişmesine yol açan başka bir yol olarak görülebilir çünkü Siber-Fiziksel Sistem sensör verilerini işlemek ve fiziksel sistemi kontrol etmek için dijital bir ikiz kullanabilir (Alam ve El Saddik 2017).

## 4. DİJİTAL İKİZ İŞLEYİŞİ

Şekil 2'de gaz halinde bulunan bir maddenin basınçlı iki adet tank içinde yer aldığı bir sistemin dijital ikiz modellemesi yer almaktadır. Şeklin sol tarafında bulunan "Fiziksel Ortam" tarafı gerçek dünyada bu tankların bulunduğu fiziksel ortamı göstermektedir. Bu ortamda tankların üzerinde çeşitli duyargalar (sensör) bulunmaktadır. Gerçek zamanlı olarak bu duyargalar vasıtası ile tanklar üzerinden bilgi toplanmakta ve bu bilgiler bir iletişim hattı üzerinden "Sanal Ortam" olarak şeklin sağ tarafında temsil edilmiş olan bilgisayar ortamında oluşturulmuş dijital kopyaya aktarılmaktadır. Bu şekilde fiziksel nesnenin, sürecin veya ortamın gerçek zamanlı veriler ile sanal bir ikizi oluşturulur. Sanal ikiz üzerinde çeşitli algoritmalar vasıtası ile makine öğrenmesi ve yapay zeka gibi yöntemler uygulanabilir hale gelmektedir. Buradan elde edilen sonuçlar ve kararlar iletişim hattı vasıtası ile fiziksel ortama iletilir. Böylece fiziksel ortamda gelen bilgiler ile fiziksel ortamda çeşitli eylemler gerçekleşmektedir.

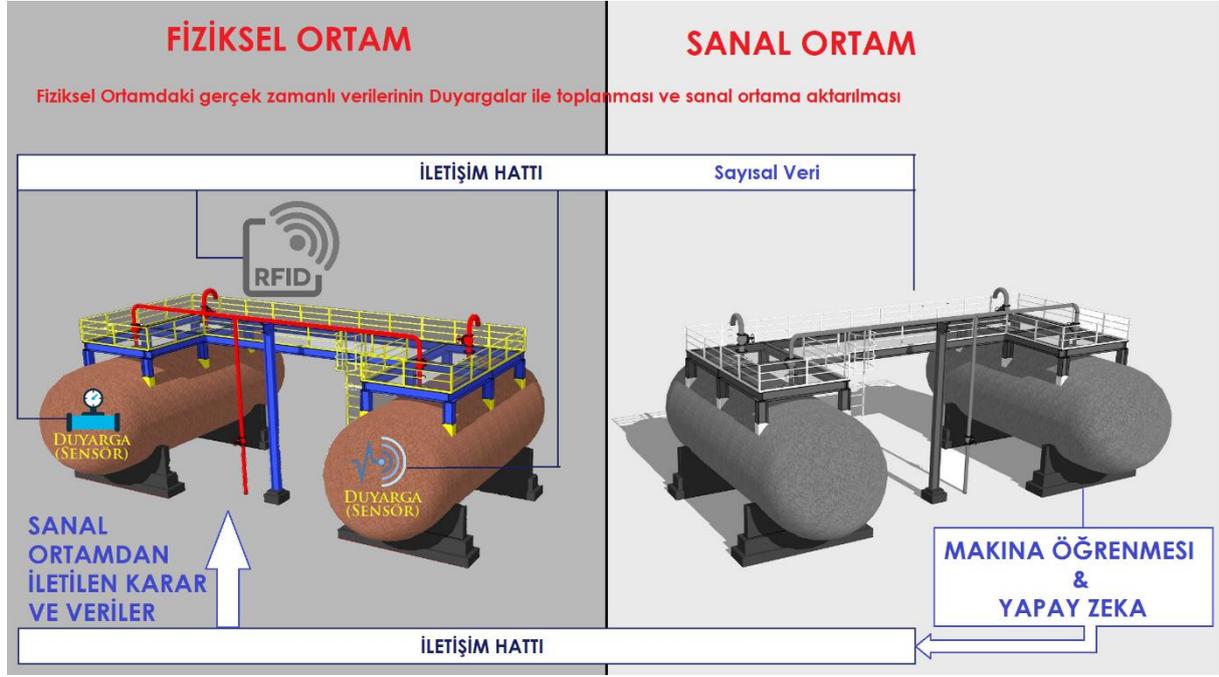

**Şekil 2. Basınçlı Tankların Digital İkiz Modeli ile Takibi ve Yönetilmesi**

Dijital ikiz kavramsal mimarisi Tablo 1'de verilmiştir. Dijital ikiz kavramsal mimarisi işlemleri 6 başlık altında gruplandırılabilir. Tablonun üst kısmında yer alan oluşturma, iletişim, toplama, analiz işlem adımları fiziksel işlem alanından dijital ortama doğru işlem aşamalarını göstermektedir. Altta yer alan anlamlandırma ve kavrama ile işlem yapma adımları ise dijital ikiz tarafından fiziksel ortama doğru gerçekleşen işlem adımlarıdır. Bu başlıklar aşağıdaki şekilde detaylandırılabilir.

**Tablo 1. Dijital İkiz Kavramsal Mimarisi (Parrott ve Warshaw 2017)**

| OLUŞTURMA | | İLETİŞİM | | TOPLAMA | ANALİZ | |
|---|---|---|---|---|---|---|
| Fiziksel İşlem Alanı | | İletişim Araçları | Dijital ikiz | | Erişim cihazları | |
| Bağlamsal veriler (Hava, sıcaklık vb.) | ERP Sistemi & CAD Modelleri & Üretim yürütme yazılımı | Duyargalar (Sensör) ve Fiziksel cihaz yönetim donanımları | Makine öğrenmesi, Yapay zeka, idrak ve karar verme ile ilgili sistemler, melez sistemler | | Uyarı sistemleri | |
| | | | | | Görüntüleme Sistemleri | |
| | | | | | Arayüzler | |
| **İŞLEM YAPMA** | | | | **ANLAMLANDIRMA ve KAVRAMA** | | |

**Oluşturma:** Bu adımda fiziksel işlemin birçok duyarga vasıtası ile donatılması ve kritik girdi verilerinin oluşturulması gerçekleşir. Duyargalar tarafından toplanan veriler sınıflandırıldığında genel olarak iki grup oluşur. Bunlardan ilki yer değiştirme, gerilme mukavemeti, renk bütünlüğü ve tork vb. gibi üretken varlığın fiziksel performans kriterlerine ilişkin operasyonel ölçümler; ikinci ise barometrik basınç, ortam sıcaklığı ve nem seviyesi gibi fiziksel bir varlığın işlemlerini etkileyen harici veya çevresel verilerdir. Bu sistemlerde yapılan analog ölçümler çeşitli cihazlar ile sayısal iletilere dönüştürülür ve bilgisayar ortamındaki dijital ikizine iletilir. Duyargalardan toplanan veriler, kurumsal kaynak planlama (ERP) yazılımları, imalat yürütme sistemleri (MES) veya CAD modelleri gibi sistemlerden gelen süreç bazlı bilgiler ile zenginleştirilebilir (Parrott ve Warshaw 2017).

**İletişim**: Bu adım fiziksel ortam ile dijital ortam arasındaki çift yönlü, kesintisiz ve gerçek zamanlı veri akışını sağlar. İletişim üç ana bağlama sahiptir (Parrott ve Warshaw 2017).

• Uçta Hesaplama: Kaynaklardan gelen verileri ve işaretleri işleyerek veri ortamına aktarır. Bu şekilde ağ iletişimindeki zorlukları ortadan kaldırır ve bazı özel protokollerin zor anlaşılan veri tiplerini daha kolay anlaşılır veri formatlarına çevirir.

• İletişim arayüzleri: Bu elamanlar ile veriler sensörlerden entegrasyon ortamına aktarılır.

• Uç güvenlik: Gelişen sensör ve iletişim teknolojileri yeni güvenlik risklerini ortaya çıkarmaktadır. Şifreleme, sertifikalar, uygulama anahtarları ve güvenlik duvarı gibi yöntemler en yaygın kullanılan yaklaşımlardır. Diğer tüm IoT bazlı sistemler gibi dijital ikizde güvenlik anlamında yeni yaklaşımlara ihtiyaç duymaktadır.

**Toplama:** Bu adımda verilerin toplanması, işleme veya analize hazır hale getirilmesi sağlanır (Parrott ve Warshaw 2017).

**Analiz:** Bu adımda elde edilmiş verilerin analizi yapılır ve görselleştirilerek sunuma hazır hale getirilir (Parrott ve Warshaw 2017).

**Anlama/Kavrama:** Bu adım ile analize dayalı, görselleştirilmiş veriler veya sonuçlar sunulur ve olası araştırma veya değişiklik ihtiyacı duyulan alanlar dijital ikiz modeli ile fiziksel ortamdaki analog ortamın uyumsuzlukları veya farklılıkları ortaya koyulur (Parrott ve Warshaw 2017).

**Eylem:** Bu adım ile birlikte önceki adımlarda elde edilen, anlamlandırılan ve karar verilen bilgilerin fiziksel ortama geri bildirim yapılması gerçekleşir (Parrott ve Warshaw 2017).

Dijital ikiz modeli ile işletmeler genellikle fiziksel nesne veya süreçlerin modellenmesinde yukarıdaki adımları kullanırlar. Büyük veri işleme teknolojileri, çok yönlü analiz teknolojileri, depolama olanaklarının büyük boyutlara ulaşması, standartlaştırılabilir verilerle bütünleşebilme gibi imkanlar ile Dijital ikiz modeli zenginleşerek uygulaması kolaylaşmıştır.

## 5. DİJİTAL İKİZ KATKISI VE KULLANIM ALANLARI

Dijital ikizin potansiyel faydalarına bakacak olursak üretim planlamasının optimize edilmesi, risklerin ve darboğazların ortaya koyulması, üretim süreçlerinin en aza indirilmesi ve kullanılan kaynakların değerlendirilmesi sayılabilir. Dijital ikiz, fiziksel ikizine kıyasla toplam yaşam döngüsü yaklaşımını sunmaktadır. Ürünlerin tasarlanması, yenilerinin tanıtımının yapılması, üretim süreçlerinin kurulması ve bütünleşik tedarik zincirinin yönetilmesi bu konsept ile birlikte kolaylaşmıştır. Ayrıca tasarım, simülasyon ve yaygın olarak karşımıza çıkan modelleme mühendisliği uygulamaları ile kıyasladığımızda bir Dijital ikizin temel farkı toplam yaşam döngüsü için gerçek ve sanal kısım arasında kalıcı ve çift taraflı bir bağlantı olmasıdır (Koulamas ve Kalogeras 2018).

Bu teknoloji ile birlikte birçok gerçek dünya zorluğunu ortadan kaldıracak çözümler ortaya koyulmaya başlamıştır. Deniz ve rüzgar tribünlerindeki yorgunluk ve korozyon direncinin test edilmesinden tutun yarış araçlarında verimlilik iyileştirilmesine yardımcı olmasına kadar geniş bir yelpazede kullanım örneklerine rastlanabilmektedir. Dijital ikiz sayesinde üretim süreçlerinin iyileştirilmesi, ürün yaşam döngüsünün genişletilmesi ve ürün geliştirme aşamalarında çeşitli çözümler bulmak ve araştırmak mümkün hale gelmiştir. Burada temel noktaların başında maliyet gelmektedir. Bilindiği gibi fiziksel bir ortamın test edilmesi veya yeniden kurulması maliyetli bir işlemdir. Dijital ikiz ile gerçek dünyada bulunan fiziksel ortamın gerçek verileri ile birlikte dijital ortamda bir ikizinin oluşturulması ile birlikte bu çözümler ve işlemler maliyet ve uygulanabilirlik açısından gerçekleştirilmesi çok kolaylaşmıştır.

Dijital ikiz, ürün yaşam döngüsünde çok fazla fayda sağlamaktadır. Dünyanın en büyük şirketlerinden Siemens (Boger ve Rusk 2017) gibi birçok şirket tarafından geliştirildi ve Gartner tarafından "Top 10 Strategic Technology Trends" listesine üç kez art arda seçildi (Panetta 2016; Panetta 2017; Panetta 2018).

Dijital ikiz kavramı hızlı bir şekilde gelişmesine rağmen gerçek dünya uygulamaları bakımından hala birçok eksikliği vardır. Bunun da temel nedeni dijital ikizin, birçok sistemin karmaşık bir dijital kopya sistemi olmasından kaynaklıdır.

Bu yöntemin kullanıldığı örneklerden birisi de sağlık alanında hastanelerdir. Dijital ikiz ile hastane, yöneticiler, doktorlar, hemşireler ve diğer tüm çalışan ve süreçler sistem üzerine dijital olarak aktarılmıştır. Dijital kopyası oluşturulan bu sistem ile hasta sağlığı ve iş akışları konusunda analizler, testler, simülasyonlar ve iyileştirmeler yapmasına olanak sağlanmıştır. Duyargalar vasıtası ile toplanan canlı veriler bu alanlarda kullanılmakta ve bu sayede hastanelerde ciddi maliyet tasarrufları ortaya koyulabilmektedir (Aynacı 2020).

Öte yandan NASA bu teknolojinin ilk kullanıldığı yerlerden birisidir. Uzay ve hava araştırmaları konusunda bu sistemler büyük katkı sunmaktadır. NASA fiziksel olarak yakınınızda olmayan başka bir deyişle uzayda bulunacak bir sistemin işletiminin, bakımının veya onarımının nasıl yapılacağı konusunda ciddi çözümlere ihtiyaç duymaktadır. NASA araştırma bölümü bu anlamda fiziksel olarak görme ve izleme yeteneğinin ötesine geçecek sistemler geliştirmek isterken karşılaştığı temel zorluk bu idi. Günümüze geldiğimizde NASA yeni öneriler, yol haritaları ve araçlar geliştirmek için dijital ikizleri kullanmaktadır (Shafto ve diğerleri 2010; Shafto ve diğerleri 2012).

Önümüzdeki beş yıl içinde milyarlarca nesne, süreç veya hizmetin dijital ikizlerle temsil edileceği öngörülmektedir. Dijital ikiz teknolojisi, şirketlerin müşteri ihtiyaçlarını daha iyi anlayarak, mevcut ürünler, operasyonlar ve hizmetler üzerinde geliştirmeler yapmalarını sağlayacaktır. Dijital ikiz, müşteri deneyimini geliştirecek ve memnuniyeti arttıracaktır.

## 6. DİJİTAL İKİZİN MALİYETE ETKİSİ

Maliyet düşürme tekniği olarak dijital ikiz kavramı basit bir şekilde şöyle bir benzetme ile anlatılabilir. Kişisel olarak kullandığını bir odanız var ve odanızın içerisine yerleştirmeniz gereken masa, sandalye(ler), koltuk, yatak, halı elbise ve kitap dolapları, sehpa, etajer gibi eşyalar var.

Eşyaları yerleştirdiniz. Çok zaman harcadınız ve yoruldunuz. Ancak yaptığınız yerleşim içinize sinmedi. Kitap dolabı çalışma masasına uzak kaldı. Yatak elbise dolabına uzak kaldı. Halı tüm zemini iyi kapatmadı, önemli bir kısmı yatağın altında kaldı.

Odayı yeniden düzenleme istiyorsunuz, bütün eşyaların yerinin değişmesi gerekiyor. Bunun için odadaki eşyaların bir kısmını dışarı çıkartıp, diğer kısmını öngördüğünüz yerine koyup tüm eşyaları yeni yerine yerleştirmeye çalışıp hoşunuza giden bir yerleşim olup olmadığına bakıp karar vermeniz gerekmektedir.

Yeni yerleşim biçimi hoşunuza giderse ne ala, gitmezse eşyaları eski yerlerine koyup eski yerleşime döneriz veya yeni düzenlemeler üreterek yapmaya çalışırız.

İdeali bulana kadar eşyaların yerini hep değiştirip olup olmadığına karar verdikten sonra olmamış ise yeni arayışlara yine eşyaların yerini değiştirerek bulmanız gerekecek. Buda hem zaman kaybı hem de yorgunluk demektir.

Odaya eşya yerleştirme işlemini odanın ve eşyaların dijital birer kopyasını bilgisayar ortamında oluşturarak yapabiliriz. Eşyaları bilgisayar ortamında odaya birçok seçeneği değerlendirerek çok sayıda ihtimal ile (belki de sonsuz) yerleştirebiliriz.

Buna ilişkin bilgisayar programını kendimiz oluşturabilir veya hazır programlardan satın alarak (Mimarların kullandığı) eşyaları odaya optimum şekilde yerleştirebiliriz.

Dijital ikizde yerleştirme karar verip uygulamaya ondan sonra geçirip zaman kazanmak, yorgunluk önleme, eşyaların yıpranmasını önleme, olası yaralanmaların önüne geçme, döşemelere vb. zarar vermenin önüne geçme gibi çeşitlendirilebilecek birçok fayda sağlayacaktır.

Siyasal, sosyal ve ekonomik krizler ve teknolojinin getirdiği yenilikler günümüz dünyasında rekabeti üst düzeye çıkarmış ve işletmelerin de bu rekabet ortamında varlıklarını koruyabilmeleri için daha düşük maliyetle yüksek kalitede ürünler üretmesi sonucunu doğurmuştur. Bu doğrultuda sermaye yatırımlarını en alt düzeyde tutarak, verimliliği ve üretkenliği eniyileme yoluna başvurmak zorunda kalmışlardır. Bunun gibi birçok nedenden dolayı yeni arayışlar içine giren işletmeler yeni yöntemler ile üretim zamanını ve maliyetleri azaltırken, verimliliği arttırmaktadırlar (Yükçü 2000, 18).

Daha önce açıkladığımız üzere dijital ikiz uygulaması günümüzde etkili bir maliyet düşürme tekniği olarak kullanılmaktadır veya kullanılmalıdır. Dijital ikizin maliyet düşürmeye etkisini açıklamadan önce "Geçiş Zamanı" (Throughout Time) kavramının üzerinde durmak yararlı olacaktır.

*Geçiş Zamanı*: Bir mamulün üretime başlamasından tüketiciye gönderilmek üzere hazır hale gelmesine kadar geçen zamana geçiş zamanı adı verilir (Yükçü 2000, 18).

Geçiş zamanı, üretim zamanı, kontrol zamanı, taşıma zamanı, bekleme zamanı ve depolama zamanı gibi zaman dilimlerinden oluşmaktadır. Bu zamanlara dikkat edilecek olursa bazı zamanlar mamulün üretimine değer katarken bazı zamanların değer katmadığı görülecektir. Buna göre (Yükçü 2000):

*Geçiş Zamanı = Değer katılan zaman + Değer katılmayan zaman* veya

*Geçiş Zamanı = Üretim Zamanı + Değer katılmayan zaman*

Bir işletmenin geçiş zamanı aşağıdaki sürelerden oluştuğunu düşünelim.

| | |
|---|---|
| Üretim Zamanı | : 5 saat |
| Kontrol Zamanı | : 3 saat |
| Taşıma zamanı | : 1.5 saat |
| Bekleme zamanı | : 6.5 saat |
| Depolama Zamanı | : 24 saat |
| Toplam | : 40 saat |

Bu aşamada yeni bir kavram (oran) ve bu kavramın hesaplanması yerinde olacaktır. Değer katma oranı üretime değer katılan oranların toplam geçiş zamanına oranlanması ile hesaplanır.

*Değer katma oranı = Değer katılan zaman / Toplam geçiş zamanı*

Bu formülü yukarıdaki verilere uygularsak değer katma oranı 5 saat / 40 saat = 0,125 olarak hesaplanır. Çünkü üretime sadece üretim zamanı boyunca değer katılmaktadır. Diğer zamanlarda üretime hiç değer katılmamaktadır. %12,5 olarak hesaplanmış olan oranın bu işletme için çok düşük olduğu görülecektir. İşletmenin hedefi değer katma oranını olabildiğince yükseltmek olmalıdır. Mümkün ise %100 'e çıkarmaktır. Ancak o zaman maliyetler düşürülebilir.

Geçiş zamanı içerisinde yer alan taşıma zamanı, bekleme zamanı ve depolama zamanı dikkat çekicidir. Bu zamanları da azaltmak (sıfıra indirme) mümkün olmalıdır. Dijital ikizin amacı da gereksiz maliyet oluşturan taşıma zamanı, bekleme zamanı ve depolama zamanı gibi zamanları dijital ortamda kurgulayıp en aza indirmektir.

Birçok işletmede değer katmayan zamanlar set-up (kurulum), kurulumun testi, malzemenin aktarılması, yerleştirilmesi, boşaltılması, yarı mamullerin depolanması, mamullerin depolanması, ara kalite kontroller ve son kalite kontroller ile ortaya çıkar. Dijital ikiz bu zaman kayıplarını en aza indirgeyecek yeni bir yaklaşımdır. Dijital ikiz yardımı ile bu bekleme süresinde ortaya çıkan aşağıdaki maliyet kalemlerinde tasarruf edilebilir.

- Direkt İşçilik Gideri
- Genel Üretim Gideri
    - o Amortisman
    - o Endirekt işçilik
    - o Enerji
    - o Kira
    - o Memur ücreti
    - o Üretime düşen vergi, harç
    - o Üretime düşen formen payı
    - o Genel Yönetim Gideri - Üretime düşen pay
    - o Pazarlama satış Dağıtım Gideri - Üretime düşen pay

Yukarıdaki maliyet kalemlerinden üretime düşen payın bekleme süresiyle heba edilen kısmını dikkate almak gerekir. Aşırı bekleme nedeni ile bozulma ve çürümenin neden olduğu direkt ilk madde ve malzeme giderleri de buna dâhil edilebilir. Bu tutarlar çok önemli maddi değerlere ulaşabilmektedir. Dijital ikiz uygulaması bu kayıpları tasarruf etmeyi sağlayacaktır.

Ambalaj üreticisi TrakRap CEO'su Martin Leeming, şirketinde herhangi bir metali kesmeden önce denenebilen, test edilebilen ve rafine edilebilen bir dijital ikiz oluşturma becerisine sahip olduklarını belirtmiştir. Ayrıca fiziksel ikizlerin çalışacağı ortamı veya fabrikayı simüle edebilir ve hatta neredeyse devreye alabilecek durumda olduklarını söylemiştir. Bu şekilde dijital ikizin maliyetlerini %50 oranında düşüreceğini beyan etmiştir (Willamson 2017).

Dijital ikiz, otomobil imalat sektöründeki üretim ve geliştirme maliyetlerini en aza indirmek için etkili bir araç olarak büyük bir başarı ile kullanılmıştır. Carlos Miskinis, 2018 yılında yaptığı çalışmasında dijital ikiz kullanımı ile otomobil imalat sektöründe üretim ve geliştirme maliyetlerinin %54 oranında azaltıldığı belirtilmiştir (Miskinis 2018).

## 7. DİJİTAL İKİZ İLE MALİYET DÜŞÜRME MODELLEMESİ

Dijital ikiz yardımı ile maliyet düşürme yaklaşımını aşağıdaki gibi bir modelle örnekleyebiliriz.

Örnek X işletmesi fiziksel olarak talaşlı imalat bölümünün makine parkını geleneksel yöntem veya dijital ikiz uygulaması ile her yıl yeniden dizayn edecektir. Dizayn aşamasında makineler yerlerinden alınarak yeni yerine temellendirilecek ve böylece malzeme akışı hızlandırılacaktır. Talaşlı imalatta çok çeşitli CNC tezgâhlar, matkaplar kullanılmaktadır. Geleneksel yöntemde her yeniden dizayn üç hafta sürmektedir. Dijital ikizde ise bu işlem 3 günde yapılabilmektedir. Bunun nedeni ise gerçek veriler ile dijital ortamda oluşturulan ikizin üzerinde uygun seçenekler denenerek optimum sonuca ulaşabilmesi, taşıma sürecinde yaşanabilecek tüm durumların öngörülebilmesidir. Bu örnekte günlük üretim kaybı işletmeye 100000 TL'ye mal olacaktır. Dijital ikiz programlama çalışması yazılımı 1000000 TL'ye mal olmakta ve sistemi çalıştıracak eleman aylık olarak işletmeye 20000 TL'lik bir maliyet getirmektedir. Sensörler vb. dijital ikiz için gerekli teknik donanım ise 600000 TL'lik bir maliyet oluşturmaktır.

Bu maliyet verileri ile iki sistemi karşılaştıracak olursak Tablo 2'deki gibi bir maliyet tablosu ortaya çıkacaktır.

5 yıllık verilerin bugünkü değeri dikkate alınarak hesaplama yapılmıştır.

**Tablo 2. Geleneksel Yöntem ile Dijital İkiz Karşılaştırmalı Maliyet Hesabı**

|  | GELENEKSEL YÖNTEM | DİJİTAL İKİZ |
|---|---|---|
| Yeniden dizayn | 10.500.000 TL<br>(3 hafta x 7 gün x 5 yıl x 100.000 TL) | 1.500.000 TL<br>(3 gün x 5 yıl x 100.000 TL) |
| Programlama yazılımı |  | 1.000.000 TL |
| Personel |  | 1.200.000 TL<br>(20.000 x 5 yıl x 12 ay) |
| Sensörler |  | 600.000 TL |
| TOPLAM | 10.500.000 TL | 4.300.000 TL |

Geleneksel yöntem yerine dijital ikiz uygulamasının oluşturacağı varsayımsal fayda 10.500.000-4.300.000 = 6.200.000 TL olarak hesaplanmıştır. Bu şekilde maliyetlerde yaklaşık olarak %59 oranında düşüş sağlanacaktır.

## 8. SONUÇ VE ÖNERİLER

Dijital ikiz günümüz teknolojilerinden nesnelerin interneti, makine öğrenmesi ve yapay zeka gibi kavramların gelişmesi ile siber fiziksel sistemlerden türetilmiş bir model olarak karşımıza çıkmaktadır. Gelişen iletişim altyapıları ve sensörler gibi donanımların uygulama maliyetlerinin düşmesi ve alanlarının genişlemesi ile dijital ikiz modeli birçok alanda uygulanmaya başlamıştır. Bu çalışma ile dijital ikiz modelinin anlaşılmasına katkı sağlayacak bilgiler, tarihsel gelişimi ve altyapısı ortaya koyulmuştur. Birçok alanı etkilediği gibi işletmelerin olası maliyet kalemlerine olan etkisi değerlendirilmiştir.

Dijital ikiz ile gerçek dünyada yer alan fiziksel nesneler veya gerçekleştirilen süreçler, işlemler dijital dünyaya taşınacaktır. Gerçek dünyanın dijital dünyada oluşturulan bu kopyası ile verilere anlık ulaşılabilecektir. Hatalar, hileleri ve sorunları hızlıca tespit edilebileceğinden güvenilir bir sistem ortaya koyulabilecektir. Makale içinde de bahsedildiği üzere maliyeti etkileyen birçok konuyu direk olarak etkileyecektir. Gerçek dünyada yapılması zor veya maliyetli işlemlerin sanal ortamda yapılması ve sonuçlarına bakılarak gerçek dünyada uygulanması mümkün olacaktır. Bu işlemlerin sanal ortamda gerçek zamanlı verilerin işlenmesi ve anlamlandırılması ile daha kolay ve düşük maliyetle yapılmasının önü açılmıştır. Çeşitli simülasyonlar yapılarak yeniliklerin fiziksel ortamda denenmeden önce dijital olarak denenmesi mümkün hale gelecektir. Yapay zeka ve makine öğrenmesi gibi yöntemlerin kullanımı ile birçok sürecin iyileştirilmesi sağlanabilecektir. Dijital ikizde somut sorunları görüp gerçek üretim alanına aktarmak daha akılcı ve az maliyetlidir. Makine öğrenmesi ve yapay zeka gibi yöntemler ile öngörüler yapılabilmekte ve böylece sadece mevcut durumun analizi değil gelecekte oluşabilecek durumların da öngörülmesi sağlanabilmektedir. Bu işletmelerin maliyetlerine ciddi katkılar sağlayabilecek bir etki ortaya çıkarmaktadır.

**YAZARLARIN BEYANI**

Bu çalışmada, Araştırma ve Yayın Etiğine uyulmuştur, çıkar çatışması bulunmamaktadır ve de finansal destek alınmamıştır.

**AUTHORS' DECLARATION**

This paper complies with Research and Publication Ethics, has no conflict of interest to declare, and has received no financial support.

**KAYNAKÇA**

Alam, K. M. ve El Saddik, A. 2017. "C2PS: A digital twin architecture reference model for the cloud-based cyber-physical systems". IEEE Access, 5, 2050–2062. doi: 10.1109/ACCESS.2017.2657006.

Autiosalo, J. 2018. "Platform for industrial internet and digital twin focused education, research, and innovation: Ilmatar the overhead Crane". IEEE World Forum on Internet of Things, WF-IoT 2018 - Proceedings, 241–244. doi: 10.1109/WF-IoT.2018.8355217.

Aynacı, İ. 2020. "Dijital İkiz ve Sağlık Uygulamaları". İzmir Katip Çelebi Üniversitesi İktisadi ve İdari Bilimler Fakültesi Dergisi, 3(1), 70-79.

Boger, T. ve Rusk, J. 2017. "What does a Siemens digital twin look like?". Siemens Product Lifecycle Management Software Inc. https://community.plm.automation.siemens.com/t5/Digital-Twin-Knowledge-Base/What-does-a-Siemens-digital-twin-look-like/ta-p/432968 (Erişim: 21.06.2018).

Brussel, H. Van, Bongaorts, L., Wyns, J., Valckenaers, P. ve Van Ginderachter, T. 1999. "A Conceptual Framework for Holonic Manufacturing : Identification of Manufacturing Holons". Journal of Manufacturing systems, 18(1), 35–52.

Erden, A. S. 2003. "Maliyet Yönetimi ve Küresel Rekabete Yönelik Maliyetleme". Muhasebe Bilim Dünyası Dergisi, 5(4), 81-95.

Främling, K., Holmström, J., Ala-Risku, T. ve Kärkkäinen, M. 2003. "Product agents for handling information about physical objects". Helsinki University of Technology Laboratory of Information Processing Science, TKO-B 153(03), 20. http://www.cs.hut.fi/Publications/Reports/B153.pdf. ISBN 951-22-6853-1 (Erişim:25.02.2020).

Grieves, M. 2005. "Product lifecycle management: the new paradigm for enterprises". International Journal of Product Development, 2(1/2), 71–84. doi: 10.1504/IJPD.2005.006669.

Grieves, M. 2011. Virtually Perfect: Driving Innovative and Lean Products through Product Lifecycle Management. Cocoa Beach, Florida: Space Coast Press, LLC. ISBN 978-0982138007.

Grieves, M. ve Vickers, J. 2017. "Digital twin: Mitigating unpredictable, undesirable emergent behavior in complex systems". Editörler: Kahlen, F. J., Flumerfelt, S., and Alves, A. Transdisciplinary Perspectives on Complex Systems: New Findings and Approaches. Springer, Cham, 85–113. doi: 10.1007/978-3-319-38756-7_4.

Holmström, J., Främling, K., Tuomi, J., Kärkkäinen, M. ve Ala-Risku, T. 2002. "Implementing Collaboration Process Networks". The International Journal of Logistics Management, 13(2), 39–50. doi: 10.1108/09574090210806414.

Hribernik, K., Rabe, L., Schumacher, J. ve Thoben, K.-D. 2005. "A concept for product- instance-centric information management". 2005 IEEE International Technology Management Conference (ICE). 20-22 June 2005. Munich, Germany: IEEE, 1–8. doi: 10.1109/ITMC.2005.7461301.

Hribernik, K. A., Rabe, L., Thoben, K. D. ve Schumacher, J. 2006. "The product avatar as a product-instance-centric information management concept". International Journal of Product Lifecycle Management, 1(4), 367–379. doi: 10.1504/IJPLM.2006.011055.

Koestler, A. 1970. "Beyond atomism and holism—the concept of the holon". Perspectives in Biology and Medicine, 13(2), 131-154.

Koulamas, C., ve Kalogeras, A. 2018. "Cyber-physical systems and digital twins in the industrial internet of things [cyber-physical systems]". Computer, 51(11), 95-98.

Kiritsis, D. 2011. "Closed-loop PLM for intelligent products in the era of the Internet of things". CAD Computer Aided Design. Elsevier Ltd, 43(5), 479–501. doi: 10.1016/j.cad.2010.03.002.

Lee, E. A. 2008. "Cyber physical systems: Design challenges". 11th IEEE International Symposium on Object and Component-Oriented Real-Time Distributed Computing (ISORC). 5-7 May 2008. Orlando, FL, USA, 363–369. doi: 10.1109/ISORC.2008.25.

Miskinis, C. 2018. "Digital Twin Genie Case Study: 54% reduction in automotive manufacturing costs". https://www.challenge.org/insights/digital-twin-genie-in-manufacturing/. (Erişim: 25.02.2020)

Panetta K. 2016. "Gartner's Top 10 Strategic Technology Trends for 2017 - Smarter With Gartner". Gartner, Inc. https://www.gartner.com/smarterwithgartner /gartners-top-10-technology-trends-2017/ (Erişim: 25.02.2020).

Panetta, K. 2017. "Gartner Top 10 Strategic Technology Trends for 2018 - Smarter With Gartner". Gartner, Inc. https://www.gartner.com/smarterwithgartner /gartner-top-10-strategic-technology-trends-for-2018/ (Erişim: 25.02.2020).

Panetta, K. 2018. "Gartner Top 10 Strategic Technology Trends for 2019 - Smarter With Gartner". Gartner, Inc. https://www.gartner.com/smarterwithgartner/gartner-top-10-strategic-technology-trends-for-2019/ (Erişim: 25.02.2020).

Parrott, A. ve Warshaw, L. 2017. "Industry 4.0 and the digital twin:Manufacturing meets its match" . Deloitte University Press. https://www2.deloitte.com/content/dam/insights/us/articles/3833_Industry4-0_digital-twin-technology/DUP_Industry-4.0_digital-twin-technology.pdf (Erişim:25.02.2020).

Shafto, M., Conroy, M., Doyle, R. ve Glaessgen, E. 2010. "DRAFT Modeling, Simulation, information Technology & Processing Roadmap". Technology Area 11. http://www.nasa.gov/pdf/501321main_TA11-MSITP-DRAFT-Nov2010-A1.pdf (Erişim:25.02.2020).

Shafto, M., Conroy, M., Doyle, R. ve Glaessgen, E. 2012. "Modeling, Simulation, information Technology & Processing Roadmap". Technology Area 11. https://www.nasa.gov/sites/default/files/501321main_TA11-ID_rev4_NRC-wTASR.pdf (Erişim:25.02.2020).

Schroeder, G. N., Steinmetz, C., Pereira, C. E. ve Espindola, D. B. 2016. "Digital Twin Data Modeling with AutomationML and a Communication Methodology for Data Exchange". IFAC-PapersOnLine. Elsevier, 49(30), 12–17. doi: 10.1016/J.IFACOL.2016.11.115.

Willamson, J. 2017. "Digital twin helps us to reduce costs by over 50%". https://www.themanufacturer.com/articles/digital-twin-helps-us-to-reduce-costs-by-over-50/. (Erişim: 25.02.2020)

Wooldridge, M. ve Jennings, N. R. 1995. "Intelligent agents: theory and practice". The Knowledge Engineering Review, 10(2), 115–152. doi: 10.1017/S0269888900008122.

Wong, C. Y., McFarlane, D., Ahmad Zaharudin, A. ve Agarwal, V. 2002. "The intelligent product driven supply chain". IEEE International Conference on Systems, Man and Cybernetics. *6-9 Oct. 2002*. Yasmine Hammamet, Tunisia, Tunisia: IEEE, 6. doi: 10.1109/ICSMC.2002.1173319.

Yükçü, S. 2000. "JIT Üretim Sisteminin Maliyet Muhasebesi Uygulamalarına Etkisi", Muhasebe ve Denetime Bakış Dergisi, 1, 18-30.